\documentclass[twocolumn,showpacs,preprintnumbers,amsmath,amssymb,nofootinbib]{revtex4}
\usepackage{amsfonts}
\usepackage{amsmath}
\usepackage{graphicx}%Include figure files
\usepackage{dcolumn}%Align table columns on decimal point
\usepackage{bm}
\usepackage{color}
\newlength\savedwidth

\newlength\savewidth

\begin{document}

\draft

\title{Immediate Causality Network of Stock Markets}
\author{Li Zhou$^1$, Lu Qiu$^2$, Changgui Gu$^1$, Huijie Yang$^1$\footnote{Corresponding author. E-mail: hjyang@ustc.edu.cn}}
\address{$^1$ Business School, University of Shanghai for Science and Technology,Shanghai 200093, China\\
$^2$ School of Finance and Business, Shanghai Normal University, No. 100 Guilin Rd., Shanghai 200234, China}
\date{\today}
\begin{abstract} \textnormal{\small {A financial system contains many elements networked by their relationships. Extensive works show that topological structure of the network stores rich information on evolutionary behaviors of the system such as early warning signals of collapses and/or crises. Existing works focus mainly on the network structure within a single stock market, while a collapse/crisis occurs in a macro-scale covering several or even all markets in the world. This mismatch of scale leads to unacceptable noise to the topological structure, and lack of information stored in relationships between different markets. In this work by using the transfer entropy we reconstruct the influential network between ten typical stock markets distributed in the world. Interesting findings include, before a financial crisis the connection strength reaches a maxima, which can act as an early warning signal of financial crises; The markets in America are mono-directionally and strongly influenced by that in Europe and act as the center; Some strongly linked pairs have also close correlations. The findings are helpful in understanding the evolution and modelling the dynamical process of the global financial system. }}\\
\\
keywords: stock market network; transfer entropy matrix
\end{abstract}

\pacs{89.75.-k, 05.40.-a, 05.45.-a}
\maketitle
\section{INTRODUCTION}

Globalization networks the financial systems distributed all over the world into a complex system. Stock markets are barometers of their corresponding local financial systems respectively, relationship network between which displays subsequently the state of the global finance \cite{nature_01,nature_02} .

Complex network theory has been adopted in some works to monitor evolutionary behaviors of financial systems\cite{topology01, topology02, topology03, topology04, topology041, topology05, topology051, topology06, topology07,topology08, topology09, topology10, topology11, topology12, topology13,topology14}. For instance, Song et al.\cite{topology03} convert cross-correlations between a total of 57 industry portfolios in the American stock market into a series of planar maximally filtered graphs (a network embedded in a high-dimensional surface) to represent the states of the corresponding successive durations, and find that mutual entropy between successive states reaches a peak at a financial crisis. Qiu et al.\cite{topology14} construct a series of stock networks from successive segments of return series covering two months each to describe the evolutionary process of the Dow Jones stock market in the duration from 1994 to 2013. Linking the stock networks between which the distances are less than a threshold, results in a state network. Distribution of the stock networks corresponding to famous crises in history on the state network illustrates specific characteristics of the crises.

Zheng et al.\cite{topology05} find that the signs of components in the eigenvectors of stock cross-correlation matrix corresponding to the few largest eigenvalues can divide stocks into two groups. The major group, defined to be the one contains more industries in the US stock market, covers however the blue chips and ST stocks in the Chinese stock market. Mantegna et al. \cite{topology07} calculate the cross-correlation matrix of a total of 49 industry indices of the American stock market portfolio. They find that the largest components in the eigenvectors corresponding to the largest two eigenvalues are closely related with the financial crises in history. Ren et al.\cite{topology09} report that the cross-correlation coefficient and the largest eigenvalue of the cross-correlation matrix of stock prices in the Shanghai Stock market increase significantly in the periods of the dot-com bubble in 2001 and the global financial crisis in 2008.

But all the works try to construct a network between stocks within a specific stock market, while a collapse and/or crisis occurs generally in a scale covering several markets or even the global financial system. This mismatch of scale implies several limitations. First, the price and/or volume for every stock is determined by the special and flexible conditions of the corresponding company, and is strongly polluted by noise, which make the extracted relationships between it and the other stocks have unacceptable low-confidence. The non-trivial patterns in the stock network are generally contaminated by statistical fluctuations. Second, the state of a specific stock market is an integrated result of interplays between social/economical/financial systems all over the world. A global crisis may induce significant changes or even strong shocks to the stock market, but the change of the stock market solely is not enough to coordinate the crisis. Third, the non-trivial information stored in the relationships between markets is lost.

Hence, in the present work we will construct a network between ten typical stock markets in the world. The stock market network is used to represent the state of the global financial system. An event in the system is then displayed by its damage to the network. Technically, we collect the indices of a total of ten stock markets distributed in Asia, America, and Europe. Let a window covering one year slide along the multivariate series with a step of one month. From every segment covered by the window, the influences of every stock market on the others are measured quantitatively by transfer entropies. The stock market networks corresponding to the successive segments form a series, which contains the evolutionary information of the global financial system.

Interesting findings include, before each famous financial crisis the influential strength between the markets reaches a maxima, which can be regarded in a certain degree as an early warning signal; though averagely the influences between the markets are symmetrical, the markets in USA are strongly and mono-directionally influenced by that in Europe and act as the center; Some strongly influential relationships have strong cross-correlations. These findings shed new lights on the coupling structure of the global financial system and are helpful in modelling the dynamical processes on it.

\section{Materials and Methods}
\subsection*{Data}

Ten stock markets distributed in different continents are considered, i.e, the Don Jones(DJI) and NASDAQ (NASD) in USA, the Tokyo's Nikkei (NIKK), Hongkong's HangSeng (HSI), Shanghai's Stock Market (SHI), Shenzhen's Stock Market (SZI), and Taiwan's TWII in Asia, and German's DAX, London's FTSE, and Paris' CAC40 (CAC) in Europe. The original data is daily indices in the duration from January 1992 to March 2017 (a total of  $5040$ simultaneous records)  \cite{stock}.

Let us denote the price series with,
\begin{flalign}
\begin{split}
P=\left\{ \begin{matrix}
   {{P}_{1,1}} & {{P}_{1,2}} & \cdots  & {{P}_{1,T+1}}  \\
   {{P}_{2,1}} & \ddots  & \cdots  & {{P}_{2,T+1}}  \\
   \vdots  & \vdots  & \ddots  & \vdots   \\
   {{P}_{M,1}} & {{P}_{M,2}} & \cdots  & {{P}_{M,T+1}}  \\
   \end{matrix} \right\}\text{  ,} \\
    \end{split}
\end{flalign}
\noindent where $P_{i,t}$ is the index of the $i$th stock market on the $t$th day, $M=10$ represents the number of the stock markets, $T+1=5040$ is the total length of the series. The daily return series reads,
\begin{equation}
\begin{array}{l}
R=\{ r_{i,t}\equiv \ln {P_{i,t+1}}-\ln {P_{i,t}},\\
i=1,2,,\cdots,M;t=1,2,\cdots,T\}.
\end{array}
\end{equation}

Based upon the top ten financial events each year reported in the \textit{Financial Times (in Chinese)}\cite{ftimes} and the discussions in references \cite{topology03,topology04,topology051,topology06,topology07} of the present paper, eight crises are identified (see Table I).

\begin{table}[ht]
\caption{Famous crises in the duration from $1992$ to $2017$}
  \begin{center}
  \begin{tabular}{ p{1cm}<{\centering} | p{2.5cm}<{\centering} | p{4cm} }
  \hline
  ID No.& Starting time (yyyy/mm/dd) & Event  \\
  \hline
  C1&1994/12/30 &  Mexico's financial crisis  \\
  \hline
  C2&1997/07/02 &  Asian financial crisis   \\
  \hline
  C3&2002/09/23& Internet bubble burst  \\
  \hline
  C4&2005/05/29& The European referendum rejected the "EU constitutional treaty",the euro suffered heavy losses\\
  \hline
  C5&2008/09/14& The Global financial crisis in 2008\\
  \hline
  C6&2009/12/08& The world's three largest Rating firms downgraded the Greece's sovereign rating\\
  \hline
  C7&2012/05/01& Japanese financial crisis\\
  \hline
  C8&2014/12/16&Russian financial crisis\\
  \hline
  \end{tabular}
  \end{center}
\end{table}

\subsection*{Series of stock market network}
Let a window with predefined length and step slide along the return series. Successive segments covered by the window form a series of segments. From each segment we calculate the transfer entropies from every stock market index to the others to measure quantitatively the stock market's influences on the others. Transfer entropies between all the stock markets form a transfer entropy matrix, called herein stock market network. By this procedure, the multivariate series is converted to a series of stock market network, which stores the evolutionary behavior of the global financial system.

Denoting the length and step of the window with $L$ and $\Delta$, the segment series reads,
\begin{flalign}
\begin{split} \label{eq:extracted_R}
& {{R}_{s}}=\left\{ \begin{matrix}
   {{r}_{1,\Delta \centerdot (s-1)+1}} & {{r}_{1,\Delta \centerdot (s-1)+2}} & \cdots  & {{r}_{1,\Delta \centerdot (s-1)+L}}  \\
   {{r}_{2,\Delta \centerdot (s-1)+1}} & \ddots  & \cdots  & {{r}_{2,\Delta \centerdot (s-1)+L}}  \\
   \vdots  & \vdots  & \ddots  & \vdots   \\
   {{r}_{M,\Delta \centerdot (s-1)+1}} & {{r}_{M,\Delta \centerdot (s-1)+2}} & \cdots  & {{r}_{M,\Delta \centerdot (s-1)+L}}  \\
\end{matrix} \right\}\text{  ,} \\
 & s=1,2,\cdots W ,\\
 \end{split}
\end{flalign}
\noindent where $W$ is the total number of segments,  $[.]$ the rounding of a real number .

The segment series is then mapped to a series of stock market network, $S_{\rm TE}(s),s=1,2,\cdots,W$, whose element $[S_{\rm TE}(s)]_{\rm mn}$ is the influence of the $m$th market on the $n$th market in the duration corresponding to the $s$th segment, measured quantitatively herein with transfer entropy. The specific form of transfer entropy proposed in the paper \cite{plosone} is employed. Let us denote the return series of the $m$th and $n$th stock markets in the $s$th segment with $X$ and $Y$, namely,
\begin{equation}
\left[\begin{matrix}
X\\
Y\\
\end{matrix} \right]
\equiv
\left[\begin{matrix}
r_{m,\Delta\cdot (s-1)+1},r_{m,\Delta\cdot (s-1)+2},\cdots,r_{m,\Delta\cdot (s-1)+L}\\
r_{n,\Delta\cdot (s-1)+1},r_{n,\Delta\cdot (s-1)+2},\cdots,r_{n,\Delta\cdot (s-1)+L}\\
\end{matrix} \right]
\end{equation}
\noindent and define two corresponding series with a delay of $\tau$,
\begin{equation}
\left[\begin{matrix}
X^\tau(l)\\
Y^\tau(l)\\
\end{matrix} \right]
\equiv
\left[\begin{matrix}
R_s(m,l-\tau)\\
R_s(n,l-\tau)\\
\end{matrix} \right],
l=\tau+1,\tau+2,\cdots,L.
\end{equation}
\noindent The transfer entropy from $X$ to $Y$ reads,
\begin{equation}
S^\tau_{\rm TE} (m,n) = H(Y|Y^1)-H(Y|Y^1,X^\tau),
\end{equation}
\noindent where $H(Y|Y^1)$ and $H(Y|Y^1,X^\tau)$ are the Shannon entropies of $Y$ under conditions of $Y^1$ and $Y^1,X^\tau$,
\begin{equation}
\begin{matrix}
H(Y|Y^1)= -\sum\limits_{y\in Y,y^1\in Y^1} P(y,y^1)lnP(y|y^1),\\
H(Y|Y^1,X^\tau)=-\sum\limits_{y\in Y,y^1\in Y^1, x^\tau\in X^\tau} P(y,y^1,x^\tau) lnP(y|y^1,x^\tau),\\
\end{matrix}
\end{equation}
\noindent respectively.

Here are two remarks. First, we must consider the time zone effect. The ten stock markets distribute in different time zones. The opening and closing times for the stock markets are different. A stock market opening later may receive information on the other markets opening earlier, and be affected subsequently by the markets. From the viewpoint of time zone, the stock markets cluster into three groups distributed in the Asia, Europe, and America respectively. Within each group the time zones for the stock markets are identical or have a slight difference, e.g., the Shanghai Stock Market in China Mainland opens an hour earlier than the NIKK in Japan. However, the difference of opening times of stock markets in different groups can reach six or even twelve hours. Hence, a proper method to consider the time zone effect is: within each group the markets can not communicate effectively in the same day; a market in the Europe can receive the information of the markets in the Asia group in the same day; and a market in the America can receive information from all the markets in the Asia and Europe.

Second, we are interested specially in the immediate influence, namely, how the present record influences the state of the next record. Because of the high-efficiency of information spreading, if we consider a large value of $\tau$, the influence will be an integration of mutual-influences within a total of $\tau$ days. The influences between a pair of stock markets will entangled together. Hence, the immediate influence is a reasonable selection. The value of $\tau$ depends subsequently on the time zones of stock markets. If the market $A$ opens several hours earlier than the market $B$, the influence of $A$ on $B$ is calculated with $\tau=0$, otherwise $\tau=1$. By this way all the markets are networked together, called immediate
causality network of stock markets.

\section{Result}

In our calculations, the window size is selected to be $L=12$-calendar-month, the sliding step $\Delta=1$-calendar-month. The return series is then separated into a total of $292$ segments. From the resulting series of $S_{\rm TE}^\tau(s),s=1,2,\cdots,W$, several properties of the networks are calculated.

\textbf{Average of influences} (AVI) between the stock markets is defined simply from summation of all the elements in the stock market network,
\begin{equation}
I_{\rm ave}(s)\equiv \frac{1}{M(M-1)}\cdot \sum\limits_{m\ne n=1}^{M} \left[S_{\rm TE}^\tau(s)\right]_{m,n}.
\end{equation}
A small (large) value of $I_{\rm ave}$ implies a weak (strong) average influence between the stock markets. We are specially interested in the behavior before or in an event of financial collapse.

\begin{figure}
\center\scalebox{0.75}[0.75]{\includegraphics{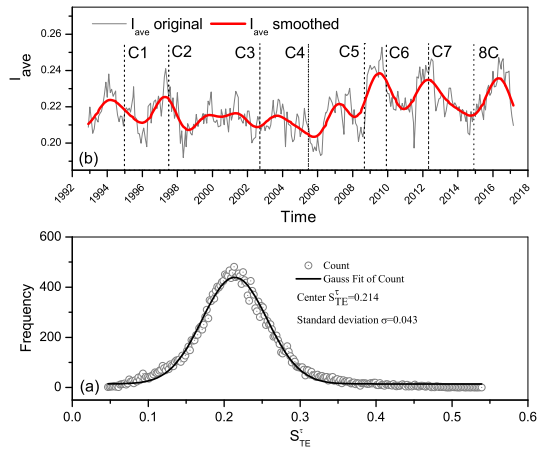}}
\caption{Average of influence as a prior indicator of financial crises. (a) distribution of transfer entropy estimated from
  all the transfer entropies. (b) Evolution of average influence. Time duration for each stock market network is represented
  with the last day of the corresponding segment of return series, namely, the last day of the $[12+(s-1)*1]$th month. The original value of $I_{\rm ave}$ (grey curve) fluctuates frequently. Filtering out high-frequent components results in the trend (red curve). The starting times of the eight famous crises are marked with vertical dotted lines respectively. Before each crisis there appears a peak, which can act as an early warning signal.} \label{Fig.1}
\end{figure}

From Fig.1(a) one can find that the transfer entropy distributes normally, centers at $0.214$ with a standard deviation of $0.043$.  Fig.1(b) presents the evolution of AVI in the years from 1992 to 2017. The last day of the $[12+(s-1)*1]$th month in the return series is used to represent the duration corresponding to the $s$th segment, namely, the value of $I_{\rm ave}(t)$ is obtained from the segment ended at $t$. $I_{\rm ave}(t)$ fluctuates frequently with $s$ (the gray curve).

A reasonable assumption is that an intrinsic behavior of the financial system should evolves slowly and is perturbed strongly by the daily changes of environments. A Fast-Fourier Transformation procedure is conducted to filter out the noises (high-frequency components) and expose subsequently the trend, as shown with the red curve. Interestingly, the trend has several peaks, several months later from each of which one can find a famous financial collapse in the world. The peaks and the crises are one-to-one correspondence, except the last financial crisis occurred in Russia in 2014. The summits of peaks corresponding to the crises $C1,C2,C3,C4,C5,C6$ and $C7$ listed in Table I occur at $1994/04,1997/04,2001/05,2003/09,2007/04,2009/09$ and $2012/05$ , namely, $8,3,16,20,17,3$ and $0$ months before the occurrences of the crises. As for the crisis $C8$ in Russia, it is well-known that it is a localized collapse with ignorable influence on the global financial system. Hence, the AVI seems to be a good indicator prior a financial collapse.

Extensive works in literature have found that a crisis occurs generally around a peak of cross-correlation. But there exist much more peaks around which one can find no crises. Hence, flittering noise (high-frequency-components) from the original curve of influence is the key procedure to find an early warning signal of financial crisis, which is ignored by almost all the researchers.

A tentative understanding of the behavior is that: When the financial system works in a formal way, people in the world tend to cooperate with each other (increase of AVI), which implies a high-efficiency and a high-risk at the same time; When a risk occurs and spreads in the financial system, people working in the system and being aware of the risk try to eliminate and/or block it by weakening the relations between different markets (a decrease of AVI); A failure of the efforts may induce a crisis.

\textbf{Asymmetry of influence} (ASI) is used to show if the influence between a pair of markets is mono-directional or bidirectional. Herein it is measured by the ratio of average difference between over average summation of transfer entropies of a pair of stock markets,
\begin{equation}
\begin{matrix}
&I_{\rm asy}(s) \equiv \frac{\sum\limits_{m>n=1}^M \left| \left[S_{\rm TE}^\tau(s)\right]_{m,n}-\left[S_{\rm TE}^\tau(s)\right]_{n,m} \right|}{\sum\limits_{m>n=1}^M \left\{\left[S_{\rm TE}^\tau(s)\right]_{m,n}+\left[S_{\rm TE}^\tau(s)\right]_{n,m}\right\}}\\
&=\frac{\sum\limits_{m>n=1}^M \left| \left[S_{\rm TE}^\tau(s)\right]_{m,n}-\left[S_{\rm TE}^\tau(s)\right]_{n,m} \right|\cdot \frac{1}{M(M-1)}}{I_{\rm ave}}.\\
\end{matrix}
\end{equation}

\begin{figure}
\center\scalebox{0.7}[0.7]{\includegraphics{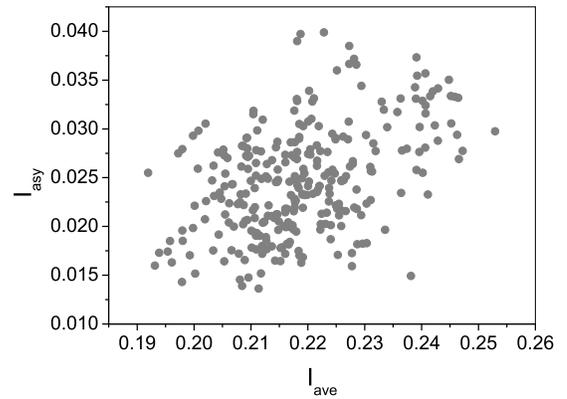}}
\caption{Asymmetrical behaviors of the influences. Statistically the influence is symmetrical.}\label{Fig.2}
\end{figure}

\noindent In the scatter plot of $\left(I_{\rm ave}, I_{\rm asy}\right)$, the points will cluster around the original point $(0,0)$ if there is almost no influence between the stock markets, i.e., the global financial system is in a loosely linked state. If the stock markets influence bidirectionally with a strong strength on each other, the points will distribute in the bottom-right corner. A point at the top-right corner implies that for each specified pair of stock markets, the influence tends to be mono-directional.

As shown in Fig.2, the values of $I_{\rm ave}(s),s=1,s,\cdots,W$ distribute in a wide range of $[0.190,0.255]$, while that for the values of $I_{\rm asy}(s),s=1,s,\cdots,W$ in a sharp interval of $[0.01,0.04]$. And the cross-correlation between $I_{\rm ave}$ and $I_{\rm asy}$ is very weak (if there exists any). Hence, influence between the stock markets is averagely symmetrical.

\textbf{Activity of influence} of the $m$th stock market on the $n$th one is evaluated by the average and the variance of transfer entropy, which are estimated respectively with,
\begin{equation}
\begin{matrix}
&A_{\rm str} (m,n) = \frac{1}{W}\cdot \sum\limits_{s=1}^W  \left[S_{\rm TE}^\tau(s)\right]_{m,n},\\
&A_{\rm flu} (m,n) = \sqrt{\frac{1}{W}\cdot \sum\limits_{s=1}^W  \left\{\left[S_{\rm TE}^\tau(s)\right]_{m,n}-\overline{\left[S_{\rm TE}^\tau(s)\right]_{m,n}} \right\}^2},\\
\end{matrix}
\end{equation}

\begin{figure}
\center\scalebox{0.35}[0.35]{\includegraphics{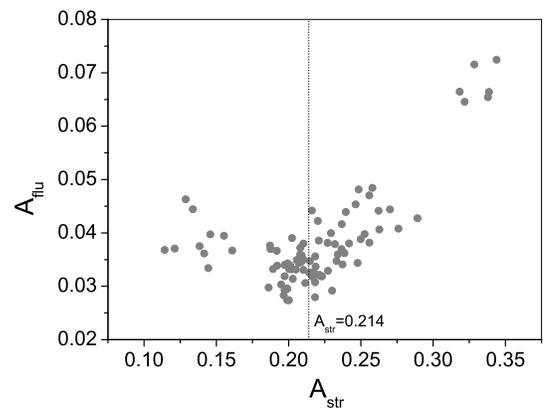}}
\caption{Activity of influence. Average of $A_{str}$ is marked with the vertical dotted line.}\label{Fig.3}
\end{figure}

\noindent where $\overline{\cdot}$ is statistical average. A large(small) value of $A_{\rm str}$ implies a strong(weak) average influence of the $m$th market on the $n$th one. A large(small) value of $A_{\rm flu}$ shows that the influence of the $m$th market on the $n$th one is unstable (stable). For a strong and persistent influence, the point of $(A_{\rm str},A_{\rm flu})$ will appear in the bottom-right corner in the scatter plot.

Fig.3 presents the points of $(A_{\rm str},A_{\rm flu})$ for all the pairs of stock markets. There exists a strong cross-correlation between the two measures; When $A_{\rm str}$ is less than the average value of $0.214$ (indicated with the vertical dotted line), $A_{\rm flu}$ is inversely proportional to it; After then $A_{\rm flu}$ is proportional to it.

\begin{figure}
\center\scalebox{0.4}[0.4]{\includegraphics{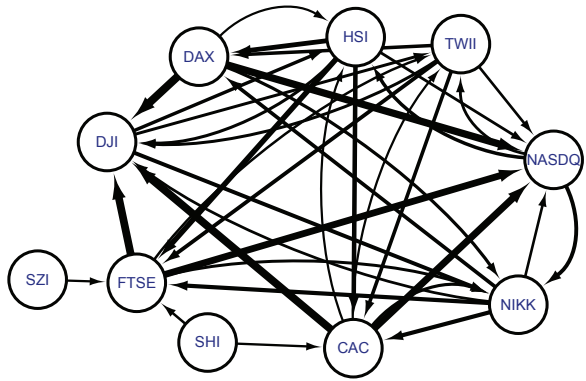}}
\caption{Influential network. When the time average of transfer entropies for a specific pair of markets is less than
the average of all the transfer entropies (the points occurring in left of the vertical dotted line in Fig.3), the
link is discarded. Strong and mono-directional influences of the markets in Europe on that in America form the main
core part of the network. Markets in America are hubs.}\label{Fig.4}
\end{figure}

\textbf{Influential network} (INN) of the stock markets are constructed with the strong influences larger than the average (occurring in the right of the vertical dotted line in Fig.3), as shown in Fig.4, namely we filter out the influences whose averages are small and fluctuations are large. One can find that the stock markets localized in America (NASDQ and DJI) are influenced strongly by, but have almost no influence on that in Europe (FTSE, CAC, and DAX). The influences between the markets in America\&Europe and that in Asia (except that in Mainland China) are bidirectional and weak. The markets in mainland China have only weak and mono-directional influences on the FTSE and CAC in Europe, namely, they can be regarded to be isolated in a certain degree. Hence, the stock markets in America are the hubs.

\begin{figure*}
\center\scalebox{0.75}[0.75]{\includegraphics{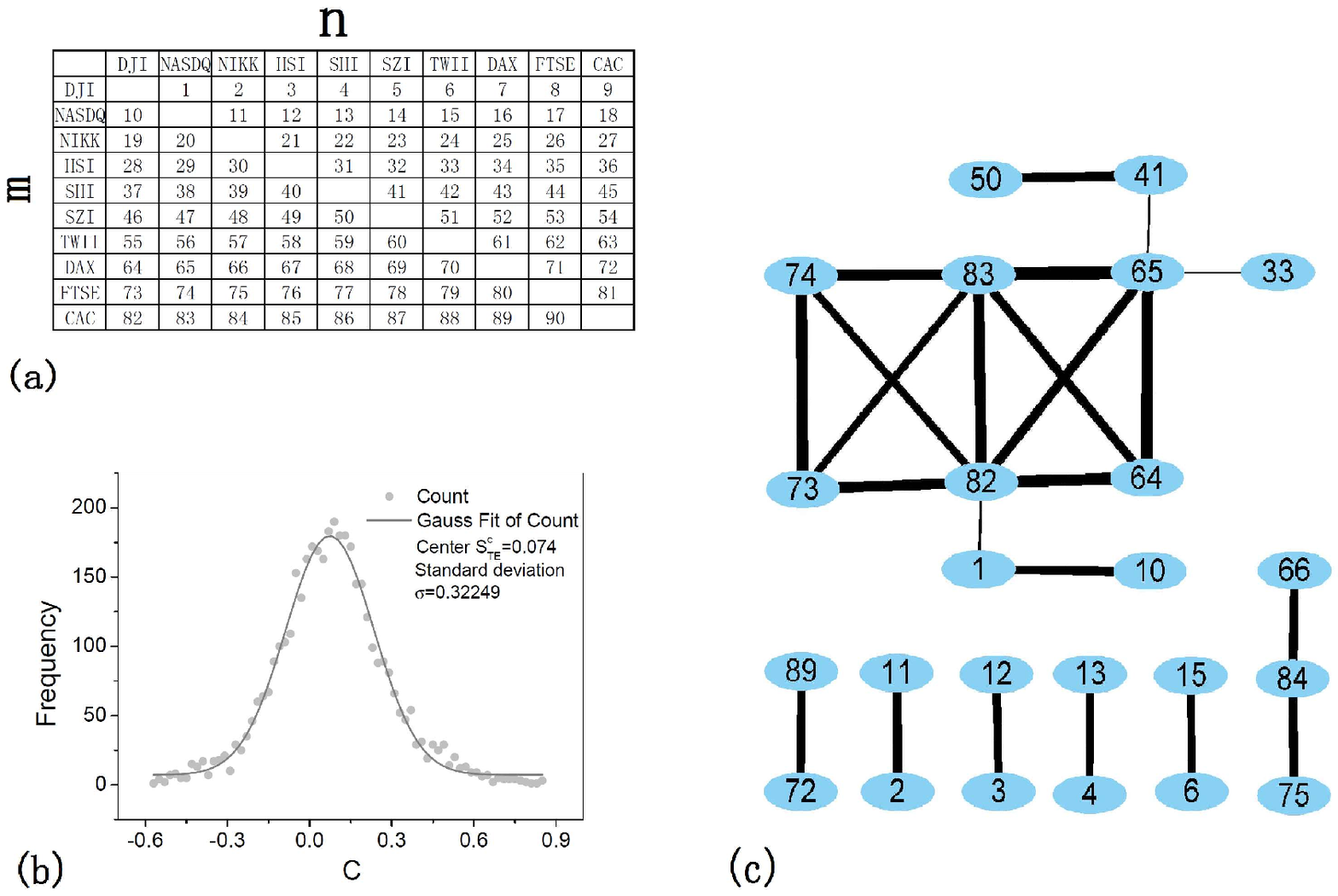}}
\caption{Influential pair network. (a) Each influential pair is assigned an identification number. The element at the row $m$
and the column $n$ is the identification number of the influential pair from the market $m$ to the market $n$. (b) The
cross-correlation coefficients between all the pairs distribute normally. (c) Discarding all the links whose absolute values of
cross-correlation coefficients are in $[0.074-0.32249,0.074+0.32249]$, namely, two standard deviations around the average value of
cross-correlation coefficients in (b), results into a influential pair network. The width of a link is proportional to the
corresponding absolute value of cross-correlation coefficient.
}\label{Fig.5}
\end{figure*}

\textbf{Influential pair network} (IPN) is then constructed to show the relations of the influential pairs. From the series of $S_{\rm TE}^\tau(s), s=1,2,\cdots,W$, one can extract two series $\left[S_{\rm TE}^\tau(s)\right]_{m,n}, s=1,2,\cdots,W$ and $\left[S_{\rm TE}^\tau(s)\right]_{i,j}, s=1,2,\cdots,W$, describing the influences of the $m$th on the $n$th and the $i$th on the $j$th market respectively. The cross-correlation coefficient of the two series, $C(m,n;i,j)$ is then calculated to show how the relation from the $i$th to $j$th stock markets on that from the $m$th to $n$th markets.

Each pair of the stock markets are assigned an identification number, as shown in Fig.5(a). The element at the row $m$ and the column $n$ is the identification number for the pair from the market $m$ to the market $n$. For instance, the identification number $47$ denotes the influential pair from SZI to NSDQ. The values of $C$ distribute normally, centering at $0.074$, with a standard deviation of $0.3225$ (see Fig.5(b)).

Here we select the standard deviation as a threshold of coupling. If the absolute value of $C(m,n;i,j)$ is larger than the threshold, we construct a link between the pair from $m$ to $n$ and the pair from $i$ to $j$. This procedure results into a influential pair network as depicted in Fig.5(c), where the width of a link is proportional to the corresponding absolute value of cross-correlation coefficient.

The strong couplings can be catalogued into three kinds. The first kind occurs between mutual pairs, i.e., the pair from $m$ to $n$ and the pair from $n$ to $m$, including the links between the nodes numbered $1$ and $10$, $41$ and $50$, and $72$ and $89$, namely between DJI$\rightarrow$NASDQ and NASDQ$\rightarrow$DJI, SHI$\rightarrow$SZI and SZI$\rightarrow$SHI, and DAX$\rightarrow$CAC and CAC$\rightarrow$DAX respectively. The second kind occurs between the pairs from stocks in Europe to that in America, including the links between $74$ and $82$, $73$ and $83$, $64$ and $83$ and $65$ and $82$. The pairs from
the markets in Europe to that in America are mediated into a cluster by the markets in America. The third kind
occurs when two pairs share one stock market, including all the other strong links ($13$ out of the total of $20$ strong links).

\section{Conclusion and discussion}

A financial system contains many elements that are linked by complicated relationships between them, which can be modelled with a network. The nodes and edges are the elements and their relationships respectively. Extensive works show that a crisis may induce significant changes to the topological structure, which in turn can be used as clue of financial collapse and/or crisis. Most of the existing works pay attentions on intra-network structure of stocks within a single financial market, while a financial collapse/crisis reaches generally several stock markets or even the whole global financial system. This mismatch of scale implies several limitations, such as large fluctuations in the network structure and the lack of important information stored in the relations between markets.

In the present work, by using transfer entropy a total of ten typical stock markets distributed in Asia, Europe, and America are linked into a series of market network, to represent the evolution of the state of the global financial system. The findings include, the linking strength between the markets provides us a significant indicator prior to financial crises; Averagely the influences between the markets are almost symmetrical, but the markets in America are strongly and mono-directionally influenced by that in Europe, and occupy the center of the system; Some influential relations closely  cross-correlated with each other; and the markets in Mainland China have limited influence on the global financial system, if there is any. These findings provide some detailed information on the coupling structure between stock markets in the world, which can be used further in modelling the dynamical processes of the system.

A system contains generally many elements that are coupled together by their relationships, which can be modelled by a complex network. Monitoring dynamical process of the network produces a multivariate time series. Reconstructing from the multivariate records the network to represent the system's state attracts special attentions in recent years. From the series of network one can obtain information on detailed structure of the system and its evolution, which is the preliminary step to develop a mathematical model of the system and has potential applications in diverse fields.

For instance, from physiological signals recorded from the brain, eyes, heart, leg, chin and lung, Ivanov et al. \cite{IVANOV} construct a series of networks between the six organisms. Link number of the network can identify different sleep stages and the transitions between the stages. From high-through output co-expressions of genes for volunteers that are initially healthy  and infected by influenza, Chen et al. \cite{CHENLN} construct a series of relation network between genes for every person. The network structure can be used to distinguish healthy individuals from the others. Using networks constructed with the fMRI/EEG/DTI records for different regions of brain, Zhang et al.\cite{ZHANGJ} monitor successfully mental disorders. Actually, our solution can be extended straightforwardly to these problems.

\section{acknowledgements}
The work is supported by the National Science Foundation of China under Grant No.10975099 (H. Yang) and No.11505114 (C. Gu), the Program for Professor of Special Appointment (Oriental Scholar) at Shanghai Institutions of Higher Learning under Grant Nos D-USST02 (H. Yang) and QD2015016(C. Gu).

% \bibliographystyle{IEEEtran}
% \bibliography{new,new2,naturetry,yahoonew}

\end{document}